\documentclass[aap,MSNbibl,citesort,dvips]{arximspdf}

\doi{10.1214/11-AAP830} 
\volume{22}
\issue{5}
\pubyear{2012}
\firstpage{2067}
\lastpage{2085}

\makeatletter
\newtheorem{thmm}{Theorem}[section]

\newtheorem{cor}[thmm]{Corollary}
\newtheorem{prop}[thmm]{Proposition}

\newproclaim{defi}[thmm]{Definition}
\newproclaim{exmp}[thmm]{Example}
\newproclaim{rem}[thmm]{Remark}
\makeatother

\begin{document}
\begin{frontmatter}

\title{Simple arbitrage}
\runtitle{Simple arbitrage}

\begin{aug}
\author{\fnms{Christian} \snm{Bender}\corref{}\ead[label=e1]{bender@math.uni-sb.de}}
\runauthor{C. Bender}
\affiliation{Saarland University}
\address{Department of Mathematics\\
Saarland University\\
P.O. Box 151150\\
D-66041 Saarbr\"ucken\\
Germany\\
\printead{e1}}
\end{aug}

\received{\smonth{6} \syear{2011}}
\revised{\smonth{11} \syear{2011}}

%
\begin{abstract}
We characterize absence of arbitrage with simple trading strategies in
a discounted market with a constant bond and several risky
assets. We show that if there is a~simple arbitrage, then there is
a~0-admissible one or an obvious one, that is, a simple
arbitrage which promises a minimal riskless gain of $\varepsilon$, if the
investor trades at all.
For continuous stock models, we provide an equivalent condition for
absence of 0-admissible simple
arbitrage in terms of a property of the fine structure of the paths,
which we call ``two-way crossing.'' This property can be verified for
many models by the law of the iterated logarithm. As an application we
show that the mixed fractional Black--Scholes model, with Hurst
parameter bigger than a half, is free of simple arbitrage on a compact
time horizon. More generally, we discuss the absence of simple
arbitrage for stochastic volatility models and local volatility models
which are perturbed by an independent 1$/$2-H\"older continuous process.
\end{abstract}

%


\end{frontmatter}

\section{Introduction}\label{sec1}

The fundamental theorem of asset pricing characterizes absence of
arbitrage in terms of the existence of equivalent martingale measures.
More precisely, the version of the fundamental theorem obtained by
Delbaen and Schachermayer \cite{DS94} states that a locally bounded
stock model does not admit a free lunch with vanishing risk,
if and only if the the model has an equivalent local martingale measure.
As absence of arbitrage is generally considered as a minimum
requirement for a sensible stock model,
nonsemimartingale models have widely been ruled out in financial
modeling. However, absence of arbitrage heavily depends
on the class of admissible strategies. In this respect the fundamental
theorem of asset prices assumes the largest possible
class of admissible strategies, namely all self-financing strategies
with wealth processes which are bounded from below.

In this paper we discuss absence of arbitrage within the class of
simple strategies. The class of simple strategies consists of
portfolios which
cannot be
rebalanced continuously in time, but only at a finite number of
stopping times. These simple strategies can actually be considered
as a reasonable description of the trading opportunities which can be
implemented in reality. Assuming a discounted model with a constant
bond and a finite number of risky assets, we first prove that if there
is a simple arbitrage (i.e., an arbitrage with a simple strategy),
then there must be one of two particularly favorable types: an obvious
arbitrage, which promises a minimum gain of some $\varepsilon$ in
those scenarios, where the investor starts to trade at all; or a
0-admissible arbitrage which can be obtained without running into debt while
waiting for the riskless gain (Theorem~\ref{thmchar1}). For models
with continuous trajectories, we further characterize the absence of
0-admissible arbitrage
in terms of a property on the fine structure of the paths which we call
two-way crossing (Proposition~\ref{prop0adm}). In the case of a single risky
asset, this property means that whenever the stock price moves from its
present level, it crosses the level immediately (i.e., infinitely often
in arbitrarily short time intervals). In the multi-asset case, this
property must hold along all measurable directions; cf. Definition
\ref{defIUD} below. We finally end with a full characterization of
absence of simple arbitrage in the case of
continuous asset prices in terms of a condition on the fine structure
of the paths (two-way crossing) and on the probability
that the asset prices stay close to their present level in the long
run; see Definition~\ref{defNOA} for a
more precise statement of this property. We also discuss how these two
properties can be checked for some mixed models, that is, for some
classical arbitrage-free model whose log-prices are perturbed by adding
some 1$/$2-H\"older continuous processes. As a particular example
we prove that the mixed fractional Black--Scholes model (a
Black--Scholes model whose log-price is perturbed by adding an
independent fractional Brownian motion) with Hurst parameter $H>1/2$ is
free of simple arbitrage. This model is known not to be
a semimartingale if $1/2<H\leq3/4$; see~\cite{Ch01}. Other model
classes, which can be shown
to have no simple arbitrage under appropriate conditions, include
mixed stochastic volatility models and mixed local volatility models.

Our results can be seen in line with some recent papers which discuss
the absence of arbitrage beyond the semimartingale setting,
by either introducing market friction, such as transaction costs (e.g.,
\cite{Gu06,GRS08,GRS10}), or by restricting the class
of admissible strategies, such as \cite{Ch03,BSV08,BSV10,JPS09}. In
particular, the articles by Cheridito \cite{Ch03} and Jarrow
et al. \cite{JPS09} are closely related. They discuss absence of
arbitrage for a subclass of simple strategies, in which, additionally,
a minimal
waiting time is imposed between two transactions. This class of
strategies is called Cheridito class in~\cite{JPS09}. Bender et al.
\cite{BSV10} show that the conditional
full support property implies the
absence of arbitrage within the Cheridito class and even in a larger
class of strategies, where the waiting time is localized
in a suitable way to include the first hitting time of a given level.
As conditional full support is easily seen to exclude obvious arbitrage\vadjust{\goodbreak}
on finite time horizon, the two-way crossing property, discussed in the
present paper, can be interpreted as a key property to extend absence
of arbitrage
from the Cheridito class to the class of all simple strategies for many models.

The paper is organized as follows. In Section~\ref{sec2} we introduce the
general setting and prove the first characterization of simple arbitrage
in terms of obvious arbitrage and 0-admissible arbitrage. Section~\ref{sec3} is
devoted to the study of 0-admissible simple arbitrage
for models with continuous paths. Several examples, including the mixed
fractional Black--Scholes model, are discussed in
Section~\ref{sec4}.\vspace*{-2pt}

\section{A characterization of simple arbitrage for right-continuous processes}\label{sec2}

In this section we provide a first characterization of simple
arbitrage. We assume that a discounted market
with $D+1$ securities is given. A constant bond $B_t=1$ and $D$ stocks
modeled by a right-continuous adapted $\mathbb{R}^D$-valued stochastic
process $X_t$, $t\in[0,\infty)$, on a filtered probability space
$(\Omega, \mathcal{F}, (\mathcal{F}_t)_{t\in[0,\infty)},P)$. The
filtered probability space is assumed to satisfy the usual conditions
of completeness and right-continuity of the filtration.

An investor can trade in the market by choosing the number of shares
held at time $t$ by a \textit{simple strategy} of the form
\[
\Phi_t=\phi_0 \mathbf{1}_{\{0\}}(t) +\sum_{j=0}^{n-1} \phi_j \mathbf{1}_{(\tau
_j,\tau_{j+1}]},
\]
where $n\in\mathbb{N}$, $0=\tau_0\leq\tau_1 \leq\cdots\leq\tau_n$
are a.s. finite stopping times with respect to $(\mathcal{F}_t)$, and
the $\phi_j$ are row vectors of $D$-dimensional, $\mathcal{F}_{\tau
_j}$-measurable random variables. Note that the trader is allowed to
trade on an infinite time horizon because we do not restrict to bounded
stopping times for the reallocation of the capital. Of course, trading
on a finite time horizon $[0,T]$ is covered by switching to the process
$(X_{t\wedge T}, \mathcal{F}_{t\wedge T})$.

As the market is already discounted, the self-financing condition on
the simple strategy $\Phi$ enforces that the investor's wealth at time
$t\in[0,\infty)$ is given by
\[
V_t(\Phi;v)=v+\sum_{j=0}^{n-1} \Phi_{\tau_{j+1}} (X_{t\wedge\tau
_{j+1}}- X_{t\wedge\tau_{j}}) ,
\]
where $v$ is the investor's initial capital. The wealth process
$V_t(\Phi;v)$ inherits right-continuity from $X$ and satisfies
\[
V_\infty(\Phi;v)=\lim_{t\rightarrow\infty} V_t(\Phi;v)=v+\sum
_{j=0}^{n-1} \Phi_{\tau_{j+1}} (X_{\tau_{j+1}}- X_{\tau_{j}}),
\]
because the stopping times $\tau_j$, $j=1,\ldots, n$, are finite
$P$-almost surely.\vspace*{-2pt}

\begin{defi}
A simple strategy $\Phi$ is:
\begin{itemize}
\item an \textit{arbitrage}, if $V_\infty(\Phi;0)\geq0$ $P$-a.s. and
$P(\{V_\infty(\Phi;0)> 0\})>0$;\vadjust{\goodbreak}
\item\textit{$c$-admissible} for some $c\geq0$, if
\[
\inf_{t\in[0,\infty)} V_t(\Phi;0) \geq-c \qquad P\mbox{-almost surely}.
\]
\end{itemize}
\end{defi}

We will speak of a simple arbitrage $\Phi$ if $\Phi$ is a simple
strategy and an arbitrage.

The two types of arbitrage are 0-admissible arbitrage and obvious
arbitrage, each of which is illustrated by one of the examples.
\begin{exmp}\label{exmparbitrage}
(i) Suppose $W_t$ a Brownian motion, and for some fixed $T>0$,
\[
X_t=\cases{
\displaystyle\frac{1}{\sqrt{2\pi(T-t)}} e^{-
{W_t^2}/{(2(T-t))}}, & \quad $0\leq t <T,$\vspace*{2pt}\cr
0, & \quad $t\geq T.$}
\]
Then, $X_t$ has continuous paths $P$-almost surely and is a local
martingale,\vspace*{1pt} which can be easily verified by an application of It\^o's
formula. As $X_0=\frac{1}{\sqrt{2\pi T}}>0$ and $X_T=0$, we observe
that the simple strategy $\Phi_t=-\mathbf{1}_{(0,T]}(t)$ is an arbitrage.

Here the arbitrage is obtained in the ``long run'' by waiting up to time~$T$.
Borrowing the terminology of Guasoni et al. \cite{GRS10} this
arbitrage is an \textit{obvious arbitrage}. This means that the arbitrage
is of the form $H \mathbf{1}_{(\sigma,\tau]}$ with $|H|=1$ almost surely,
and if the investor trades at all, that is, on the set $\{\sigma<\tau\}
$, she can be sure to have a riskless gain of at least a given constant
$\varepsilon>0$ (here: $\frac{1}{\sqrt{2\pi T}}$); compare Definition~\ref
{defNOA} below.
Note that in the present example, there is no $c\geq0$ such that the
arbitrage is $c$-admissible, thanks to
the local martingale property of $X$. Notice that a related example of
a local martingale which admits simple arbitrage
has already been given in \cite{DS94b}.

(ii) Suppose $X_t=\exp\{W_t+ t^\alpha\}$ for some $\alpha
<1/2$. By the law of the iterated logarithm, we have
\[
\inf\{t>0;  \log(X_t)>0\}=0<\inf\{t>0;  \log(X_t)<0\}=:\tau.
\]
Hence, for sufficiently large $N$, the stopping times
\[
\tau_N:=\tau \wedge  1/N
\]
satisfy $P(\{\tau_N<\tau\})>0$. As
\[
P(\{X_{\tau_N}>1\})=P\bigl(\{ W_{1/N} \neq-(1/N)^\alpha\}\cap\{\tau_N<\tau\}
\bigr)=P(\{\tau_N<\tau\})>0
\]
and $X_{\tau_N}=1$ on $\{\tau_N=\tau\}$,
the strategy $\Phi_t=\mathbf{1}_{(0,\tau_N]}$ is a simple arbitrage with
wealth process
\begin{eqnarray*}
V_t(\Phi;0)
&=& X_{t\wedge\tau_N}-X_0\\
& \rightarrow&\cases{
\exp\{W_{1/N}+(1/N)^\alpha\}-1, &\quad $\tau_N<\tau,$\vspace*{2pt}\cr
0, &\quad $\tau_N=\tau$}
\qquad (t\rightarrow\infty)\vadjust{\goodbreak}
\end{eqnarray*}
for sufficiently large $N$. Here, the arbitrage can be obtained by
trading at arbitrarily short time intervals. Moreover, it is
0-admissible, because $X_t-X_0\geq0 $ on $[0,\tau]$.
\end{exmp}

The two types of arbitrages, which were illustrated in the previous
example, are particularly favorable for an
investor: Obvious arbitrages which guarantee a minimum riskless gain if
the investor starts to trade at all; 0-admissible arbitrages which can
be obtained without running into debt while waiting for the riskless
gain.

The main result of this section shows that if there is a simple
arbitrage, then there must be one of these two favorable types.

Before we state and prove the result, we first introduce the notion of
no obvious arbitrage on an infinite time horizon. The definition is in
the spirit of Guasoni et al. \cite{GRS10}.
\begin{defi}\label{defNOA}
$X$ satisfies \textit{no obvious arbitrage} (NOA) if for every stopping
time $\sigma$ and for every $\varepsilon>0$ we have: If $P(\{\sigma<\infty
\})>0$ and $H$ is a $D$-dimensional row vector of $\mathcal{F}_\sigma
$-measurable random variables such that $|H|=1$ $P$-almost surely, then
%
\begin{equation}\label{NOA}
P\Bigl(\{\sigma<\infty\}\cap\Bigl\{\sup_{t\in[\sigma,\infty)} H(X_t-X_\sigma
)<\varepsilon\Bigr\}\Bigr)>0.
\end{equation}
\end{defi}

\begin{rem}\label{remNOA}
(i) We think of $H$ as an $\mathcal{F}_\sigma$-measurable ``direction''
(and will call an $H$ with the above properties \textit{$\mathcal
{F}_\sigma$-measurable direction} from now on).
Then, (\ref{NOA}) means that, starting from $X_\sigma$ at time $\sigma
$, along each direction the probability that the stocks do not increase
by more than $\varepsilon$ is positive. Note that by passing from $H$ to
$-H$, we also get
\[
P\Bigl(\{\sigma<\infty\}\cap\Bigl\{\inf_{t\in[\sigma,\infty)} H(X_t-X_\sigma
)>-\varepsilon\Bigr\}\Bigr)>0.
\]
Hence, along each direction the probability that the stocks do not
decrease by more than $\varepsilon$ is also positive.

(ii) In the case of a single stock $D=1$, it is clearly sufficient to
check (NOA) along the directions $+1$ and $-1$. In this case, (\ref{NOA})
simplifies to
%
\begin{equation}\label{RLC1}
P\Bigl(\{\sigma<\infty\}\cap\Bigl\{\inf_{t\in[\sigma,\infty)} X_t> X_\sigma
-\varepsilon\Bigr\}\Bigr)>0
\end{equation}
and
%
\begin{equation}\label{RLC2}
P\Bigl(\{\sigma<\infty\}\cap\Bigl\{\sup_{t\in[\sigma,\infty)} X_t< X_\sigma
+\varepsilon\Bigr\}\Bigr)>0.
\end{equation}
Condition (\ref{RLC1}) was introduced by Bayraktar and Sayit \cite
{BS10} in their study of simple arbitrage in the case
of a single stock modeled by a nonnegative, strict local martingale.

(iii) In the general case $D>1$, it is not sufficient to check (NOA)
along rational directions. Here is a simple example with two stocks:
\[
X^1_t=W_{t \wedge1},\qquad  X^2_t=UW_{t \wedge1}+(t \wedge1),
\]
where $W$ is a Brownian motion, and $U$ is uniformly distributed on
$[0,1]$ and independent of $W$. Given any stopping time $\sigma$ of the
filtration $\mathcal{F}_t=\sigma(U,   W_s,  0\leq s \leq(t\wedge
1))$ and $H=(q_1, q_2) \in\mathbb{Q}^2$, we get
\[
H(X_t-X_\sigma)=(q_1+q_2U)(W_{t \wedge1}-W_{\sigma\wedge1}) + q_2(t
\wedge1)-q_2 (\sigma\wedge1),\qquad  t\geq\sigma.
\]
As $(q_1+q_2U)\neq0$ $P$-almost surely, condition (\ref{NOA}) is
clearly satisfied along rational directions. However,
choosing $\sigma=0$ and $\tilde H=(-U,1)$, we have
\[
\tilde H(X_t-X_\sigma)= t \wedge1,
\]
which shows that (NOA) is violated along the direction $\tilde
H/|\tilde H|$.
\end{rem}

The next straightforward proposition explains how to obtain an obvious
arbitrage, if (NOA) is violated.
The simple idea is to buy $H$ shares of the stocks at time $\sigma$ and
wait until
the stock prices have increased by some $\varepsilon$ in direction $H$.
This will happen with probability 1 if (NOA) is violated at time $\sigma
$ in direction $H$.

\begin{prop}\label{propOA}
If $X$ is right-continuous and does not satisfy (NOA), then~$X$ has a
simple arbitrage.
\end{prop}

\begin{pf}
We suppose that (NOA) is violated, that is, there is a stopping time~$\sigma$,
an $\varepsilon>0$ and an $\mathcal{F}_\sigma$-measurable
direction $H$ such that $P(\{\sigma<\infty\})>0$ and
\[
P\Bigl(\{\sigma<\infty\}\cap\Bigl\{\sup_{t\in[\sigma,\infty)} H(X_t- X_\sigma
)<\varepsilon\Bigr\}\Bigr)=0.
\]
We fix a sufficiently large $K$ such that $P(\{\sigma\leq K\})>0$ and
define the stopping time $\rho:=\inf\{t\geq\sigma;   H(X_t- X_\sigma
)>\varepsilon/2\}$, which is a.s. finite
on the set $\{\sigma\leq K\}$. Then, with $\tau:=\rho\mathbf{1}_{\{\sigma
\leq K\}} + K \mathbf{1}_{\{\sigma>K\}}$
and $\tilde H=H\mathbf{1}_{\{\sigma\leq K\}} + (1,0,\ldots,0) \mathbf{1}_{\{
\sigma> K\}} $, $\tilde H \mathbf{1}_{(\sigma\wedge K,\tau]}$ is a
simple arbitrage. Indeed, $V_\infty(\tilde H\mathbf{1}_{(\sigma\wedge
K,\tau]})=H(X_\rho- X_\sigma)\geq\varepsilon/2$
on $\{\sigma\wedge K<\tau\}$, and $V_\infty(\tilde H\mathbf{1}_{(\sigma
\wedge K,\tau]})=0$ on $\{\sigma\wedge K=\tau\}$. So this arbitrage is
obvious in the terminology of Example~\ref{exmparbitrage}(i).
\end{pf}

The following theorem is our first characterization of simple
arbitrage, which is valid for right-continuous stock models.
\begin{thmm}\label{thmchar1}
Suppose $X$ has right-continuous paths. Then the following assertions
are equivalent:
\begin{longlist}[(ii)]
\item[(i)] $X$ is free of arbitrage with simple strategies.

\item[(ii)] $X$ satisfies (NOA), and $X$ has no 0-admissible arbitrage of
the form $H \mathbf{1}_{(\sigma,\tau]}$ with bounded stopping times
$\sigma\!\leq\!\tau$ and an \mbox{$\mathcal{F}_\sigma$-measurable direction~$H$.}\vadjust{\goodbreak}
\end{longlist}
\end{thmm}

As a preparation we prove two propositions which are interesting in
their own rights.

\begin{prop}\label{propRLC}
Suppose $X$ has right-continuous paths. If (NOA) holds, then every
simple arbitrage is 0-admissible.
\end{prop}

\begin{pf}
Here, the main idea is the following: If there is an arbitrage which is
not 0-admissible, then the value of the strategy will, at some time,
drop below some negative level, say $-\delta$, with positive
probability. However then the wealth process must eventually increase
by at least
$\delta$ again because it must end with a nonnegative value (due to
the arbitrage property). This turns out to be in conflict with the
(NOA) property.

In more detail, suppose $\Phi_t=\phi_0 \mathbf{1}_{\{0\}}(t) +\sum
_{j=0}^{n-1} \phi_j \mathbf{1}_{(\tau_j,\tau_{j+1}]}$ is a simple
arbitrage which is not zero admissible. We define
\[
j_0=\max\Bigl\{j=0,\ldots, n-1;  P\Bigl(\inf_{t\in[\tau_j, \tau
_{j+1})} V_t(\Phi;0)<0\Bigr)>0\Bigr\}.
\]
Setting $\tau:=\tau_{j_0+1}$, we observe that $V_\tau(\Phi;0)\geq0$
$P$-almost surely. Moreover, there is a $\delta>0$ such that
%
\begin{equation}\label{eqdeltaset}
P\Bigl(\inf_{t\in[\tau_{j_0}, \tau)} V_t(\Phi;0)\leq-2\delta\Bigr)>0.
\end{equation}
Define a stopping time $\rho$ by
\[
\rho=\inf\{t>\tau_{j_0};  V_t(\Phi;0)\leq-\delta\}\wedge\tau.
\]
By right-continuity of $X$ [and hence $V(\Phi;0)$], we have $V_\rho(\Phi
;0)\leq-\delta$ on $\{\rho<\tau\}$. The latter set has positive
probability by (\ref{eqdeltaset}).
We now choose $M$ sufficiently large such that
%
\begin{equation}\label{eqphi>0}
P(\{\rho<\tau\}\cap\{0<|\phi_{j_0}|\leq M\})>0.
\end{equation}
If this probability were not positive for sufficiently large $M$, then
$P(\phi_{j_0}=0|\rho<\tau)=1$, which contradicts $V_\rho(\Phi;0)<0\leq
V_\tau(\Phi;0)$ on $\{\rho<\tau\}$.
We now define $A:=\{\rho<\tau\}\cap\{0<|\phi_{j_0}|\leq M\}\in\mathcal
{F}_\rho$ and
\[
H(\omega)=\cases{
\phi_{j_0}(\omega)/|\phi_{j_0}(\omega)|, &\quad $\omega\in
A,$\vspace*{2pt}\cr
(1,0,\ldots,0), &\quad $\omega\notin A.$}
\]
Then, on $A$,
\[
\delta\leq V_\tau(\Phi;0)-V_\rho(\Phi;0)= \phi_{j_0} (X_\tau-X_\rho
)\leq MH(X_\tau-X_\rho).
\]
Consequently,
\[
P\Bigl(A\cap\Bigl\{\sup_{t\in[\rho,\infty)} H(X_t -X_\rho)< \delta/M\Bigr\}\Bigr) \leq
P\bigl(A\cap\{ H(X_\tau-X_\rho)< \delta/M\}\bigr)=0.
\]
Defining the stopping time
\[
\sigma(\omega)=\cases{
\rho(\omega), &\quad $\omega\in A,$\vspace*{2pt}\cr
\infty, &\quad $\omega\notin
A$,}\vadjust{\goodbreak}
\]
we get
\[
P\Bigl(\{\sigma<\infty\}\cap\Bigl\{\sup_{t\in[\sigma,\infty)} H(X_t-X_\sigma)<
\delta/M\Bigr\}\Bigr)=0
\]
in contradiction to the definition of (NOA).
\end{pf}

\begin{prop}\label{prop0adm}
Suppose $X$ is right-continuous. If $X$ has a 0-admissible simple
arbitrage, then it has a 0-admissible arbitrage of the form $H \mathbf{1}_{(\sigma,\tau]}$ with bounded stopping times $\sigma\leq\tau$ and
an $\mathcal{F}_\sigma$-measurable direction $H$.
\end{prop}

In particular, this proposition shows that the study of 0-admissible
arbitrage can be restricted to
bounded random time intervals.
\begin{pf}
Suppose $\Phi_t=\phi_0 \mathbf{1}_{\{0\}}(t) +\sum_{j=0}^{n-1} \phi_j
\mathbf{1}_{(\tau_j,\tau_{j+1}]}$ is a 0-admissible simple arbitrage. We define
\[
j_0=\max\bigl\{j=0,\ldots, n-1;  P\bigl( V_{\tau_j}(\Phi;0)=0
\bigr)=1\bigr\}.
\]
We consider the strategy $\bar\Phi_t=\phi_{j_0} \mathbf{1}_{(\tau
_{j_0},\tau_{j_0 +1}]}$. As $P( V_{\tau_{j_0}}(\Phi;0)=0)=1$, we obtain
\[
V_t(\bar\Phi;0)=\cases{
0, &\quad $t\leq\tau_{j_0},$\vspace*{2pt}\cr
V_{t}(\Phi;0), & \quad $\tau_{j_0}<
t\leq\tau_{j_0+1},$\vspace*{2pt}\cr
V_{\tau_{j_0+1}}(\Phi;0), & \quad $t> \tau_{j_0+1}.$}
\]
The value process of $\bar\Phi$ cannot drop below zero because it
coincides with zero or with the value process of the 0-admissible
strategy $\Phi$. Moreover, it is an arbitrage because $P(V_{\tau
_{j_0}+1}(\Phi;0)>0)>0$ by the definition of $j_0$. We now define
\[
\tau=\cases{
\tau_{j_0+1}, & \quad $\phi_{j_0}\neq0,$\vspace*{2pt}\cr
\tau_{j_0}, &\quad $\mbox{otherwise}$}
\]
and
\[
H=\cases{
\phi_{j_0}/|\phi_{j_0}|, &\quad $\tau>\tau_{j_0},$\vspace*{2pt}\cr
(1,0,\ldots,0), & \quad$\mbox{otherwise}.$}
\]
Then $V_t(H \mathbf{1}_{(\tau_{j_0},\tau]};0)=|\phi_{j_0}|V_t(\bar\Phi
;0)$, which immediately implies that $H \mathbf{1}_{(\tau_{j_0},\tau]}$ is
a zero-admissible arbitrage, too.

If $\tau$ is bounded, the assertion of the proposition is proved.
Otherwise, we now consider the strategies $H_K \mathbf{1}_{(\tau
_{j_0}\wedge K,\tau\wedge K]}$ for $K\in\mathbb{N}$, where
$H_K=H\mathbf{1}_{\{\tau_{j_0}\leq K\}}+(1,0,\ldots,0)\mathbf{1}_{\{\tau
_{j_0}> K\}}$.
Then
\[
V_t\bigl(H_K \mathbf{1}_{(\tau_{j_0}\wedge K,\tau\wedge K]};0\bigr)=H(X_{\tau\wedge
K\wedge t}-X_{\tau_{j_0}\wedge K\wedge t})=V_{t\wedge K}\bigl(H\mathbf{1}_{(\tau
_{j_0},\tau]};0\bigr).
\]
Consequently, $H_K\mathbf{1}_{(\tau_{j_0}\wedge K,\tau\wedge K]}$ is
0-admissible. As
\[
\bigl\{V_\infty\bigl(H\mathbf{1}_{(\tau_{j_0},\tau]};0\bigr)>0\bigr\}\cap\{\tau\leq K\}
\uparrow\bigl\{V_\infty\bigl(H\mathbf{1}_{(\tau_{j_0},\tau]};0\bigr)>0\bigr\}\qquad
(K\uparrow\infty),
\]
we get
\[
P\bigl(\bigl\{V_\infty\bigl(H\mathbf{1}_{(\tau_{j_0},\tau]};0\bigr)>0\bigr\}\cap\{\tau\leq K\}\bigr)>0\vadjust{\goodbreak}
\]
for sufficiently large $K$. Now, $V_\infty(H\mathbf{1}_{(\tau_{j_0},\tau
]};0)=V_\infty(H_K\mathbf{1}_{(\tau_{j_0}\wedge K,\tau\wedge K]};0)$ on $\{
\tau\leq K\}$, which implies that
\[
P\bigl(\bigl\{V_\infty\bigl(H_K\mathbf{1}_{(\tau_{j_0}\wedge K,\tau\wedge K]};0\bigr)>0\bigr\}\cap
\{\tau\leq K\}\bigr)>0.
\]
Thanks to the 0-admissibility of $H_K\mathbf{1}_{(\tau_{j_0}\wedge K,\tau
\wedge K]}$, we conclude that this strategy is an arbitrage.
\end{pf}

With these propositions at hand, the proof of Theorem~\ref{thmchar1}
is immediate:
\begin{pf*}{Proof of Theorem \protect\ref{thmchar1}}
(ii) $\Rightarrow$ (i) immediately follows from Propositions~\ref
{propRLC} and~\ref{prop0adm}.

(i) $\Rightarrow$ (ii): It suffices to show that (NOA) is a necessary
condition for absence of simple arbitrage, which is the assertion of
Proposition~\ref{propOA}.
\end{pf*}

As a corollary we obtain a multidimensional and infinite time horizon
version of a result by Bayraktar and Sayit \cite{BS10} for local martingales.

\begin{cor}\label{corerhan}
Suppose $X$ is right-continuous, and there is a probability measure
$Q$ equivalent to $P$ such that $X$ is a $Q$-local martingale. Then the
following assertions are equivalent:
\begin{longlist}[(ii)]
\item[(i)] $X$ has no simple arbitrage.
\item[(ii)] $X$ satisfies (NOA).
\end{longlist}
\end{cor}
\begin{pf}
In view of Theorem~\ref{thmchar1} it suffices to show that the
existence of an equivalent local martingale measure rules out the
existence of a 0-admissible arbitrage of the form $H\mathbf{1}_{(\sigma
,\tau]}$ with bounded stopping times \mbox{$\sigma\leq\tau$} and $\mathcal
{F}_\sigma$-measurable directions $H$. This follows from a routine
application of the optional sampling theorem applied to the
$Q$-supermartingale $V_t(H\mathbf{1}_{(\sigma,\tau]};0)$, which is
justified by the boundedness of $\tau$.
\end{pf}
\begin{rem}
In the setting of the previous corollary, absence of simple arbitrage
cannot be deduced directly
from the existence of an equivalent local martingale measure. As we do
not require that the wealth process of a
simple strategy is bounded from below, simple arbitrage is possible
under local martingale dynamics as illustrated in Example \ref
{exmparbitrage}(i), even on a finite time horizon. Moreover, we
emphasize that Corollary~\ref{corerhan} covers the infinite time
horizon case because we allow trading at unbounded stopping times.
\end{rem}

\section{\texorpdfstring{A\hspace*{2pt} characterization\hspace*{2pt} of\hspace*{2pt} simple\hspace*{2pt} arbitrage\hspace*{2pt} for\hspace*{2pt} continuous\hspace*{2pt} processes}
{A characterization of simple arbitrage for continuous processes}}\label{sec3}
Throughout this section we assume that the stock model $X$ has
continuous paths. Under this assumption we will characterize the
absence of 0-admissible simple arbitrage. In this way we will achieve a
second characterization of simple arbitrage in terms of the concept of
``two-way crossing,'' which we introduce next.

\begin{defi}\label{defIUD}
Suppose $\sigma$ is an a.s. finite stopping time, and $H$ is an
$\mathcal{F}_\sigma$-measurable direction. Let
\[
\sigma_{H}=\inf\{t\geq\sigma,  H(X_t-X_\sigma)>0\} .
\]

\begin{longlist}[(ii)]
\item[(i)] $X$ satisfies \textit{two-way crossing at $\sigma$ along direction
$H$} if
%
\begin{equation}\label{IUD}
\sigma_{H}=\sigma_{-H} \qquad  P\mbox{-a.s.}
\end{equation}

\item[(ii)] $X$ satisfies \textit{two-way crossing} (TWC) at bounded stopping
times (at a.s. finite stopping times) if it satisfies two-way crossing
at every bounded (a.s. finite) stopping time $\sigma$ in every $\mathcal
{F}_\sigma$-measurable direction $H$.
\end{longlist}
\end{defi}

\begin{rem}
(i) (TWC) is a condition on the fine structure of the paths. Whenever
the stock price moves from $X_\sigma$ along direction $H$, $HX_t$ will
cross the level $HX_\sigma$ infinitely often in time intervals of
length $\varepsilon$ for every $\varepsilon>0$.\vspace*{-6pt}
\begin{longlist}[(iii)]
\item[(ii)] It is obvious that in the case of a single stock $D=1$, (TWC) must
only be checked in direction $H=1$.

\item[(iii)] In the multi-asset case,
it is not sufficient to check (TWC) along rational directions. The
same counterexample as in Remark~\ref{remNOA}(iii), applies.
\end{longlist}
\end{rem}

\begin{prop}\label{propIUD}
Suppose $X$ is continuous. Then the following assertions are equivalent:
\begin{longlist}[(iii)]
\item[(i)] $X$ satisfies (TWC) at a.s. finite stopping times.

\item[(ii)] $X$
satisfies (TWC) at bounded stopping times.

\item[(iii)] $X$ has no
$0$-admissible arbitrage of the form $H \mathbf{1}_{(\sigma,\tau]}$ with
bounded stopping times $\sigma$ and $\tau$ and $\mathcal{F}_\sigma
$-measurable direction $H$.

\item[(iv)] $X$ has no $0$-admissible simple arbitrage.
\end{longlist}
\end{prop}
\begin{pf}
We first introduce the notation
%
\begin{equation}
\sigma_{H,n}=\inf\{t\geq\sigma,  H(X_t- X_\sigma)\geq1/n\}\label{eqsigma+n}
\end{equation}
for $n\in\mathbb{N}$, and note that $\sigma_{H,n}\downarrow\sigma_H$
$P$-almost surely as $n\rightarrow\infty.$

(i) $\Rightarrow$ (ii): Obvious.

(ii) $\Rightarrow$ (iii): Suppose a strategy of the form $H\mathbf{1}_{(\sigma,\tau]}$ with a.s.
finite stopping times $\sigma\leq\tau$ is an arbitrage. Of course, we
can and shall assume $P(\{\tau>\sigma\})>0$
because otherwise $V_\infty(H\mathbf{1}_{(\sigma,\tau]} ;0)=0$ $P$-almost surely.

We first consider the case $P(\{\sigma_{-H}=\sigma\}|\{\tau>\sigma\}
)=1$: Then $\sigma_{-H,n} \downarrow\sigma$ on $\{\tau>\sigma\}$ and
thus $\tau_n:=\tau\wedge\sigma_{-H,n} \downarrow\sigma$ $P$-a.s.
Hence, $P(\{\sigma< \tau_n<\tau\})>0$ for sufficiently large $n$. For
such an $\tau_n$ we have, on $\{\sigma< \tau_n<\tau\}$,
\[
V_{\tau_n}\bigl(H\mathbf{1}_{(\sigma,\tau]};0\bigr)=H(X_{\tau\wedge\tau_n}-
X_{\sigma\wedge\tau_n})=H(X_{\sigma_{-H,n}}- X_{\sigma})= - 1/n.
\]
Thus, $H\mathbf{1}_{(\sigma,\tau]}$ is not 0-admissible. Note that in this
first case we did not assume boundedness of $\sigma$ and $\tau$, and
did not not apply (TWC).

Now suppose that $P(\{\sigma^-=\sigma\}|\{\tau>\sigma\})<1$ and $\sigma
$ is bounded.
We observe that, thanks to (TWC) at the bounded stopping time $\sigma$
and the continuous
paths of $X$, $X_t=X_\sigma$ on $(\sigma,\sigma_{-H}]$ and, hence,
$V_{t}(H\mathbf{1}_{(\sigma,\tau]};0)=H(X_{\tau\wedge t}- X_{\sigma\wedge
t})=0$ for $t\in[0,\sigma_{-H}]$. If $H\mathbf{1}_{(\sigma,\tau]}$ is a
0-admissible arbitrage, then so is $H\mathbf{1}_{(\sigma_{-H}\wedge\tau
,\tau]}$. However, $(\sigma_{-H})_{-H}=\sigma_{-H}$, and so the first
case applies.

(iii) $\Rightarrow$ (iv): Proposition~\ref{prop0adm}.

(iv) $\Rightarrow$ (i): Here the idea is as follows: If (TWC) is
violated, then there is a~portfolio, whose value goes up before going down.
A 0-admissible arbitrage can be obtained by buying this portfolio today
and selling it once it has increased by some $\varepsilon$, or
else when its price returns to the current level.

Precisely, suppose that $X$ does not satisfy (TWC) at some a.s. finite
stopping time $\sigma$ in direction $H$. By passing to $-H$, if
necessary, we can assume without loss of generality that the set $A=\{
\omega, \sigma_{H}(\omega)<\sigma_{-H}(\omega)\}$ has strictly
positive probability. Note that $A\in\mathcal{F}_{\sigma_H}$. We define
the sequence of stopping times
\[
\tau_{n}= (\sigma_{-H}\wedge\sigma_{H,n}) \mathbf{1}_A  +  \sigma_H
\mathbf{1}_{A^c}.
\]
Then, $\tau_n\geq\sigma_H$ a.s. and $\tau_n> \sigma_H$ on $A$.
By construction and continuity of $X$, we have $H(X_t- X_{\sigma})\geq
0$ for $t\in(\sigma_H,\tau_n]$. Therefore the strategies $H\mathbf{1}_{(\sigma_H,\tau_n]}$, $n\in\mathbb{N}$, are $0$-admissible. As
$\sigma_{H,n} \downarrow\sigma_H$ $P$-a.s., we get $\tau_n \downarrow
\sigma_H$ $P$-a.s. Therefore,
\[
P(\{\sigma_H<\tau_n<\sigma_{-H}\} ) = P(A\cap\{\tau_n<\sigma_{-H}\})>0
\]
for sufficiently large $n$. However, on $\{\sigma_H<\tau_n<\sigma_{-H}\}$,
\[
V_\infty\bigl(H\mathbf{1}_{(\sigma_H,\tau_n]};0\bigr)=H(X_{\tau_n}-X_{\sigma_H})
=H(X_{\sigma_{H,n}}-X_\sigma)= 1/n.
\]
Consequently, $H\mathbf{1}_{(\sigma_H,\tau_n]}$ is a 0-admissible
arbitrage for suffciently large $n$.~%
\end{pf}

A combination of the previous proposition with Theorem~\ref{thmchar1}
yields the following characterization of simple arbitrage for
continuous stock models.

\begin{thmm}\label{thmchar2}
Suppose $X$ is continuous. Then, the following assertions are
equivalent:
\begin{longlist}[(ii)]
\item[(i)] $X$ does not admit a simple arbitrage.

\item[(ii)] $X$ satisfies (TWC) at bounded stopping times and (NOA).
\end{longlist}
\end{thmm}

We now briefly discuss the two-way crossing property (TWC).
It follows from Lemma V.46.1 in \cite{RW94} that (TWC) holds for
one-dimensional regular diffusions. Moreover, it is
a direct consequence of Proposition~\ref{propIUD} above that every
local martingale satisfies (TWC), because local martingale models
are clearly free of 0-admissible arbitrage. We now provide a sufficient
condition for (TWC) for mixed models, that is, models of type $M_t+Y_t$
where $M$ is a~local martingale, and $Y$ is possibly a
nonsemimartingale. The key assumption is that
the quadratic variation of the local martingale is sufficiently large
in order to compensate for the path irregularity of $Y$.
\begin{thmm}\label{thmILC}
Suppose $X_t=M_t+Y_t$, where $M$ is a $D$-dimensional continuous
$(\mathcal{F}_{t})$-local martingale, and $Y_t$ is a
$D$-dimensional $(\mathcal{F}_{t})$-adapted process.
We assume that:

(1) For every $K\in\mathbb{N}$, there is a strictly positive random
variable $\varepsilon_K$ such that for every $0\leq s\leq t \leq K$,
\[
\langle M \rangle_t-\langle M \rangle_s \geq\varepsilon_K (t-s) \mathbb{I}_D,
\]
where $\mathbb{I}_D$ is the unit matrix in $\mathbb{R}^D$;

(2) $Y$ is $1/2$-H\"older continuous on compacts, that is, for every
$K\in\mathbb{N}$, there is a positive random variable $C_K$ such that
for every $0\leq s\leq t \leq K$,
\[
|Y_t-Y_s|:=\sqrt{\sum_{d=1}^D |Y_t^d-Y^d_s|^2}\leq C_K|t-s|^{1/2}.
\]
Then, $X$ satisfies (TWC) at bounded stopping times.
\end{thmm}

\begin{pf}
We fix an arbitrary stopping time $\sigma$, which is bounded by some
$K\in\mathbb{N}$, and an $\mathcal{F}_\sigma$-measurable direction
$H$. Considering
the real valued process
\[
Z_t=HX_{\sigma+t}=HM_{\sigma+t}+HY_{\sigma+t}, \qquad 0\leq t \leq1,
\]
with respect to the filtration $\mathcal{G}_t=\mathcal{F}_{\sigma+t}$,
it is sufficient to show that
%
\begin{equation}\label{hilf01}
\inf\{t\geq0,  Z_t-Z_0>0\}=0.
\end{equation}
Indeed, this implies $\sigma_H=\sigma$ and, replacing $H$ by $-H$,
$\sigma_{-H}=\sigma$.

In order to show (\ref{hilf01}), we introduce the process $M^{H,\sigma
}_t=HM_{\sigma+t}-HM_{\sigma}$, $0\leq t \leq1$, which is an $\mathcal
{G}_t$-local
martingale with quadratic variation
\begin{eqnarray}
\langle M^{H,\sigma}\rangle_t-\langle M^{H,\sigma}\rangle_s=H(\langle
M\rangle_{\sigma+t} -\langle M\rangle_{\sigma+s})H'\geq\varepsilon_{K+1}
(t-s),
\nonumber
\\
\eqntext{0\leq s\leq t \leq1,}
\end{eqnarray}
by assumption (1). In particular, $\langle M^{H,\sigma}\rangle_t$ is
strictly increasing on $[0,1]$. We extend $M^{H,\sigma}$
to a local martingale on $[0,\infty)$ with strictly increasing
quadratic variation which satifies $\langle M^{H,\sigma}\rangle
_t\rightarrow\infty$ as
$t\rightarrow\infty$, for example, by setting $M^{H,\sigma
}_t=M^{H,\sigma}_{1}+\tilde W_t-\tilde W_{1}$ for $t\geq1$, where
$\tilde W$ is a Brownian motion. Denoting by
\[
T(t)=\inf\{s\geq0,  \langle M^{H,\sigma}\rangle_s=t\}
\]
the inverse of $\langle M^{H,\sigma}\rangle$, the
Dambis--Dubins--Schwarz Theorem (see Karatzas and Shreve \cite{KS},
Theorem 3.4.6) yields that
the process $W_t=M^{H,\sigma}_{T(t)}$ is an $(\mathcal{G}_{T(t)})_{t\in
[0,\infty)}$-Brownian motion.
By the law of the iterated logarithm (see, e.g., Theorem 2.9.23 in\vadjust{\goodbreak} \cite
{KS}) applied to $W_t$ there is a set $\Omega'$
of full $P$-measure such that for every $\omega\in\Omega'$ there is a
sequence $t_n\downarrow0$ satisfying
\[
\lim_{n\rightarrow\infty} \frac{W_{t_n}(\omega)}{\sqrt{2 t_n(\omega)
\log\log(1/t_n(\omega))}}=1.
\]
We define $s_n=T(t_n)$ and notice that $s_n\downarrow0$ and
$t_n=\langle M^{H,\sigma}\rangle_{s_n}$, because the quadratic
variation of $M^{H,\sigma}$ is strictly increasing. For suffciently
large $n\geq N_0(\omega)$, we then
obtain, on $\Omega'$,
\begin{eqnarray*}
Z_{s_n}-Z_0&=&M^{H,\sigma}_{s_n}+H(Y_{\sigma+s_n}-Y_{\sigma
})=W_{t_n}+H(Y_{\sigma+s_n}-Y_{\sigma}) \\
&\geq& \frac{1}{2} \sqrt{2 \langle M^{H,\sigma}\rangle_{s_n} \log\log
(1/\langle M^{H,\sigma}\rangle_{s_n} )} - |Y_{\sigma+s_n}-Y_{\sigma}|
\\ &\geq&
\Biggl(\sqrt{\frac{\varepsilon_{K+1}}{2}} \sqrt{\log\log(1/\langle
M^{H,\sigma}\rangle_{s_n} )} - C_{K+1}\Biggr)\sqrt{s_n}.
\end{eqnarray*}
As the right-hand side is strictly positive for sufficiently large $n$
(depending on $\omega\in\Omega'$), we get (\ref{hilf01}), and the
proof is finished.
\end{pf}

\section{Examples}\label{sec4}

We finally present some examples of models which are free of simple arbitrage,
although they may fail to be semimartingales. The models, which we
discuss here,
can be considered as mixed models in the sense that some well-known
arbitrage-free semimartingale models are combined with some H\"older
continuous processes such as fractional Brownian motion.

Throughout the section we shall work on finite time horizons. To
simplify the terminology we say that a model $(X_t,\mathcal{F}_t)$ is
\textit{free of simple arbitrage on finite time horizons} if for every
$T>0$, the model $(X_{t\wedge T},\mathcal{F}_{t\wedge T})$ has no
simple arbitrage. In view of Theorem~\ref{thmchar2} it is
straightforward to deduce:
\begin{cor}\label{corcharacterization}
Suppose $X$ is continuous. Then the following assertions are
equivalent:
\begin{longlist}[(ii)]
\item[(i)] $X$ is free of simple arbitrage on finite time horizons.

\item[(ii)]
$X$ satisfies (TWC) at bounded stopping times and (NOA) holds on
$[0,T]$ for every $T>0$; that is,
For every $[0,T]$-valued stopping time $\sigma$ and for every $\varepsilon
>0$ we have the following: If $P(\{\sigma<T\})>0$, and $H$ is an
$\mathcal{F}_\sigma$-measurable direction, then
%
\begin{equation}\label{NOAT}
P\Bigl(\{\sigma<T\}\cap\Bigl\{\sup_{t\in[\sigma,T]} H(X_t-X_\sigma)<\varepsilon\Bigr\}\Bigr)>0.
\end{equation}
\end{longlist}
\end{cor}

\subsection{Mixed Black--Scholes models}

Our first class of examples concerns ``mixed Black--Scholes models,''
that is, the log-prices of a multidimensional Black--Scholes model are
perturbed by adding H\"older continuous processes.

\begin{thmm}\label{corBmmodels}
Suppose $(W_t,\mathcal{F}_t)$ is an $N$-dimensional Brownian motion,
and~$Z_t$ is a $D$-dimensional $(\mathcal{F}_t)$-adapted process
independent of $W$, which is \mbox{$\alpha$-H\"older} continuous on compacts\vadjust{\goodbreak}
for some $\alpha>1/2$.
Further assume that the matrix $\sigma\sigma^*$ is strictly positive
definite, where $\sigma=(\sigma_{d,\nu})_{d=1,\ldots,D,  \nu=1,\ldots
,N}$. Define~$D$ stocks by
\[
X^d_t=x^d_0 \exp\Biggl\{\sum_{\nu=1}^N \sigma_{d,\nu} W^{\nu}_t
+Z^d_t\Biggr\}
\]
with initial values $x^d_0>0$ for $d=1,\ldots,D$. Then the
$D$-dimensional mixed Black--Scholes model $X_t=(X^1_t,\ldots,X^D_t)^*$
is free of simple arbitrage on finite time horizons.
\end{thmm}

\begin{pf}
In view of Corollary~\ref{corcharacterization} we have to show (TWC)
at bounded stopping times and (NOA) on $[0,T]$ for $T>0$. In order to
verify (TWC) we are going to check
the assumptions of Theorem~\ref{thmILC}. As each component~$Z^d_t$ is
$\alpha$-H\"older continuous for some $\alpha>1/2$, we can conclude
that $Z^d$ has zero quadratic variation. Applying
It\^o's formula (for Dirichlet processes), we hence obtain
$X^d_t=M^d_t+Y^d_t$ with
\begin{eqnarray*}
M^d_t&=&x^d_0+\sum_{\nu=1}^N \int_0^t \sigma_{d,\nu} X^d_s \,dW^\nu_s, \\
Y^d_t&=&\frac{1}{2} \sum_{\nu=1}^N \int_0^t \sigma_{d,\nu}^2 X^d_s
\,ds+\int_0^t X^d_s \,dZ^d_s.
\end{eqnarray*}
Here, the last integral exists as Young--Stieltjes integral (see \cite
{DN98}), because $X^d$ is $\beta$-H\"older continuous on compacts for every
$\beta<1/2$, and $Z^d$ is $\alpha$-H\"older continuous for some $\alpha
>1/2$. It is then an easy consequence of the Young--Love inequality (Theorem
2.1 in \cite{DN98}) that $\int_0^t X^d_s \,dZ^d_s$ inherits the $\alpha
$-H\"older continuity on compacts of the integrator $Z^d$. In
particular, $Y$ satisfies the
H\"older condition (2) in Theorem~\ref{thmILC}.

Now notice that the cross-variation of the components of $M$ is given by
\[
\langle M^d, M^q \rangle_t=\int_0^t X^q_s (\sigma\sigma^*)_{q,d} X^d_s
\,ds,\qquad  d,q=1,\ldots,D.
\]
As $\sigma\sigma^*\geq\varepsilon\mathbb{I}_D$ for some constant
$\varepsilon>0$, we derive
\[
\langle M \rangle_t -\langle M \rangle_s \geq(t-s) \Bigl(\varepsilon\min
_{d=1,\ldots,D} \inf_{s\in[0,K]} |X^d_s|^2\Bigr) \mathbb{I}_D
\]
for $0\leq s\leq t \leq K$. Hence condition (1) of Theorem \ref
{thmILC} is satisfied as well. Applying this theorem we get (TWC) at
bounded stopping times.

It remains to check the no obvious arbitrage condition on $[0,T]$ for
$T>0$. To this end we fix $T>0$, a $[0,T]$-valued stopping time $\sigma$
with $P(\{\sigma<T\})>0$ and an
$\mathcal{F}_\sigma$-measurable direction $H$. Notice that, due to the
independence of $W$ and $Z$, $W_{\sigma+t}-W_{\sigma}$ is a Brownian
motion independent of
$\mathcal{F}_\sigma\vee\mathcal{F}^Z$, where $\mathcal{F}^Z$ is the
$\sigma$-field generated by the process $(Z_t, t\geq0)$. Applying the
full support property of this
Brownian motion and recalling that $\sigma\sigma^*$ is positive
definite, we get for every $\varepsilon>0$,
\begin{eqnarray*}
&& P\Bigl(\sup_{t\in[0,T-\sigma]} H(X_{\sigma+t}-X_\sigma)<\varepsilon\big|\mathcal
{F}_\sigma\vee\mathcal{F}^Z\Bigr)\\
&&\qquad\geq P\Biggl(\sup_{t\in[0,T-\sigma]} \sum_{d=1}^D |H_d X^d_\sigma|\\
&&\qquad\phantom{\geq P\Biggl(}  \times\Biggl|\exp\Biggl\{\sum_{\nu=1}^N \sigma_{d,\nu}(W^\nu_{\sigma+t}-W^\nu
_{\sigma})
+(Z^d_{\sigma+t}-Z^d_\sigma) \Biggr\}-1\Biggr|<\varepsilon\Big|\mathcal{F}_\sigma\vee
\mathcal{F}^Z\Biggr)\\
&&\qquad> 0
\end{eqnarray*}
$P$-almost surely. This immediately implies
\[
P\Bigl(\{\sigma<T\}\cap\Bigl\{\sup_{t\in[\sigma,T]} H(X_t-X_\sigma)<\varepsilon\Bigr\}\Bigr)>0.
\]
Hence, (NOA) holds on $[0,T]$.
\end{pf}

\begin{rem}
(i) In the univariate case $D=1$ the H\"older condition on~$Z$ can be
weakened to 1$/$2-H\"older continuous on compacts in the previous
theorem. Indeed, in this case
it is straightforward that (TWC) for $X$ is equivalent to (TWC) for
$\log(X)$. However, (TWC) for $\log(X)$ then is an immediate consequence
of Theorem~\ref{thmILC}.

(ii) Theorem~\ref{corBmmodels} does not hold if $Z$ is only H\"older
continuous with exponent $\alpha<1/2$. A simple counterexample
in the one-dimensional case is $X_t=\exp\{W_t+t^\alpha\}$. For $\alpha
<1/2$, this model admits a 0-admissible simple arbitrage; see Example
\ref{exmparbitrage}(ii).
For $\alpha\geq1/2$, this model is free of simple arbitrage by (i);
see also
Delbaen und Schachermayer \cite{DS95} or Jarrow et al. \cite{JPS09}.
The former paper also contains a construction
of an arbitrage for
$\alpha=1/2$ in the larger class of strategies with continuous
readjustment of the portfolio. This arbitrage satisfies the
usual admissibility condition which requires that the wealth process of
the portfolio is bounded from below. For $\alpha>1/2$, such arbitrage
cannot exist
because the model has an equivalent martingale measure.
\end{rem}

\begin{exmp}[(Mixed fractional Black--Scholes model)]
A \textit{fractional Brownian motion} $Z$ with Hurst parameter $H\in
(0,1)$ is a centered Gaussian process with covariance
\[
E[Z_tZ_s]=\tfrac{1}{2} (t^{2H}+s^{2H}-|t-s|^{2H}), \qquad  t,s\geq0.
\]
By the Kolmogorov--Centsov criterion (e.g., \cite{KS}, Theorem 2.2.8),
$Z$ can be chosen $(H-\varepsilon)$-H\"older continuous
on compacts for every $\varepsilon>0$. In particular $Z$ can be chosen
$\alpha$-H\"older continuous on compacts for some\vadjust{\goodbreak} $\alpha>1/2$,
whenever $H>1/2$.
The mixed fractional Black--Scholes model is of the form
\[
X_t=x_0\exp\{\sigma W_t+ \eta Z_t + \nu t + \mu t^{2H}\}
\]
for constants $\sigma,\eta,x_0>0$ and $\mu,\nu\in\mathbb{R}$, where
$W$ is a Brownian motion, and~$Z$ is a fractional Brownian motion with
Hurst parameter $H>1/2$ independent of $W$. An application of Theorem
\ref{corBmmodels} shows that the mixed fractional Black--Scholes model
with $H>1/2$ does not admit simple arbitrage
on finite time horizons. Note that $X$ is not a semimartingale with
respect to its own augmented filtration if $1/2<H\leq3/4$, but is
equivalent to the Black--Scholes model for $H>3/4$; see, for example,
Cheridito \cite{Ch01}.
Theorem~\ref{corBmmodels} also implies that a multi-asset version of
the mixed fractional Black--Scholes model has no simple arbitrage
on finite time horizons, provided $\sigma\sigma^*$ is positive definite.
\end{exmp}

\subsection{Mixed stochastic volatility models}

We now discuss the absence of simple arbitrage for stochastic
volatility models. In order to simplify the presentation, we only treat
the case of
a single risky asset.
\begin{thmm}\label{thmstochasticvol}
Suppose $(W,B)$ is a two-dimensional Brownian motion with respect to
the filtration $(\mathcal{F}_t)$, and $Z$ and $V$ are $(\mathcal
{F}_t)$-adapted processes such that~$V$ is continuous, and $Z$ is
1$/$2-H\"older continuous on compacts. Assume that $W$ is independent of
$(B,V,Z)$. Then, for $-1<\rho<1$ and $f,g \in\mathcal{C}([0,\infty
)\times\mathbb{R})$ such that $g(t,V_t)$ is strictly positive,
\begin{eqnarray*}
X_t&=&X_0\exp\biggl\{\int_0^t f(s,V_s)\,ds +\rho\int_0^t g(s,V_s) \,dB_s \\
&&\hspace*{5pt}\phantom{X_0\exp\biggl\{}{}+ \sqrt
{1-\rho^2} \int_0^t g(s,V_s) \,dW_s + Z_t\biggr\}
\end{eqnarray*}
is free of simple arbitrage on finite time horizons with respect to the
augmentation of the filtration $(\mathcal{F}^X_t)$ generated by $X$.
\end{thmm}

\begin{pf}
In the single asset case, simple arbitrage is easily seen to be stable
with respect to composition with stricty increasing functions. Hence it
suffices to show the assertion for $\log(X_t)$. By Theorem 3.1 in
Pakkanen~\cite{Pa09},
$\log(X_t)$ satisfies conditional full support on compact time
intervals with respect to the augmentation of $(\mathcal{F}^X_t)$.
However, it is a straightforward consequence Lemma 2.9 in Guasoni et
al. \cite{GRS08} that conditional full support on $[0,T]$
implies (NOA) on $[0,T]$.
In view of Corollary~\ref{corcharacterization}, it is now sufficient
to prove that $(\log(X_t),\mathcal{F}_t)$ satisfies (TWC). We decompose
$\log(X_t)=M_t+Y_t$ with
\begin{eqnarray*}
M_t&=&\log(X_0)+\rho\int_0^t g(s,V_s) \,dB_s + \sqrt{1-\rho^2} \int_0^t
g(s,V_s) \,dW_s, \\
Y_t&=&Z_t+\int_0^t f(s,V_s)\,ds.
\end{eqnarray*}
Then $M$ is a local martingale with quadratic variation $\langle M
\rangle_t=\int_0^t g^2(s,V_s)\,ds$, and along each path $\inf_{s\in
[0,K]} g^2(s,V_s)$ is strictly
positive for every \mbox{$K>0$}. Moreover, $Y$ is 1$/$2-H\"older continuous on
compacts. Therefore $M_t+Y_t$ satisfies (TWC) thanks to Theorem~\ref{thmILC}.
\end{pf}

\begin{exmp}[(A mixed Heston model)] In the Heston model \cite{He93}, the
discounted stock price $S_t$ has the dynamics
\begin{eqnarray*}
S_t&=&S_0\exp\biggl\{\mu t-\frac{1}{2}\int_0^t V_s \,ds + \rho\int_0^t \sqrt
{V_s}\,dB_s + \sqrt{1-\rho^2} \int_0^t \sqrt{V_s} \,dW_s\biggr\}, \\
V_t&=& V_0+\int_0^t \kappa(\theta-V_s)\,ds + \sigma\int_0^t \sqrt{V_t} \,dB_s,
\end{eqnarray*}
where $(W,B)$ is a two-dimensional Brownian motion, $-1<\rho<1$, $\mu$
is the drift of the discounted stock, $\theta>0$ is the long-term limit
of the volatility, $\kappa>0$ is the mean reversion speed of the
volatility and $\sigma>0$ is the volatility of
volatility. We assume the positivity condition $2\kappa\theta\geq
\sigma^2$ which ensures the strict positivity of $V_t$.
We now define a mixed fractional version of the Heston model by
\[
X_t=S_te^{Z_t},
\]
where $Z$ is a fractional Brownian motion with Hurst parameter $H>1/2$
(adapted to some filtration with respect to which $(W,B)$
is a two-dimensional Brownian motion) independent of $W$. Then, by the
previous theorem,
$X_t$ does not admit simple arbitrage on finite time horizons with
respect to the augmentation of the filtration $(\mathcal{F}^X_t)$.
Of course, the fractional Brownian motion can be replaced by any other
1$/$2-H\"older continuous processes
independent of $W$. Moreover, mixed versions of many other stochastic
volatility can be cast in the framework
of Theorem~\ref{thmstochasticvol}
in a similar way as we demonstrated for the Heston model. These include
classical stochastic
volatility models such as the Hull--White model, the Stein--Stein model and
the Wiggins model (see \cite{MR05}, Chapter~7.4 for more details), but also
the model by Comte and Renault \cite{CR}, where volatility is driven by
a fractional Brownian motion and exhibits long memory effects. See also
the discussion in Section~\ref{sec4} of Pakkanen~\cite{Pa09} in the context of
conditional full support.
\end{exmp}

\subsection{Mixed local volatility models}

Local volatility models were introduced by Dupire \cite{Du94} in order
to capture the smile effect. Again, we will focus on the case of a
single stock
$S$ and recall that its price is governed by an SDE
\[
dS_t=\mu(t,S_t)dt+\sigma(t,S_t)\,dW_t, \qquad S_0=s_0,
\]
where $W$ is a Brownian motion. Note that the drift $\mu$ and the
volatility $\sigma$ depend on time $t$ and the spot price $S_t$. More
generally, we will now consider models, where $\mu$ and $\sigma$ may
depend on the whole past of the stock price, that is,
%
\begin{equation}\label{lvSDE}
dS_t=\mu(t,S)dt+\sigma(t,S)\,dW_t, \qquad  S_0=s_0,
\end{equation}
where $\mu,\sigma\dvtx [0,\infty)\times\mathcal{C}([0,\infty))\rightarrow
\mathbb{R}$ are progressive functions satisfying
\[
|\mu(t,x)|\leq\bar\mu x(t),\qquad \bar\sigma^{-1} x(t)\leq|\sigma
(t,x)|\leq\bar\sigma x(t)
\]
for some constants $\bar\mu>0$ and $\bar\sigma>0$ for every $t\in
[0,\infty)$ and every $x\in\mathcal{C}([0,\infty))$ with $x(0)=s_0$.
We shall assume that the SDE (\ref{lvSDE}) has a weak solution. It is
shown by Pakkanen \cite{Pa09}, Section 4.2,
that $\log(S_t)$ has conditional full support on $[0,T]$ for every
$T>0$ with respect to
the filtration $(\mathcal{F}_t^{(S,W)})$ generated by $S$ and $W$. We
now suppose that a stochastic process $Z$
independent of $(S,W)$ is given, which is $1/2$-H\"older continuous on
compacts. As stock model we now consider
\[
X_t=S_te^{Z_t}.
\]
Making use of the independence of $Z$ and $(S,W)$, the conditional full
support property can be transferred from $\log(S_t)$ to $\log(X_t)$ by
conditioning additionally on the $\sigma$-field generated by $Z$.
Hence we can again conclude that $\log(X_t)$ satisfies (NOA) on $[0,T]$
for every $T>0$ with respect to its own
augmented filtration. Moreover, by Theorem~\ref{thmILC}, it is
straightforward to verify that $\log(X_t)$ satisfies (TWC) with respect to
the augmented filtration generated by $(S,W,Z)$ and hence also with
respect to the augmented filtration generated by $X$.
Appealing to Corollary~\ref{corcharacterization} we have thus proved
the following result:
\begin{thmm}
Suppose $X_t=S_te^{Z_t}$, where $S$ is given by (\ref{lvSDE}), and $Z$
is independent of $(S,W)$ and $1/2$-H\"older continuous on compacts.
Then, $X_t$ is free of simple arbitrage on finite time horizons with
respect to the augmentation of the filtration $(\mathcal{F}^X_t)$
generated by $X$.
\end{thmm}

\section*{Acknowledgment}
The paper benefited from the helpful comments of an anonymous referee.


%


\printaddresses

\end{document}